\documentclass{article}
\usepackage{spconf,amsmath,graphicx,amsfonts,amssymb}

\usepackage{booktabs}
\usepackage{caption}
\usepackage{hyperref}
\usepackage{tablefootnote}

\usepackage{diagbox} %
\usepackage[numbers, sort&compress]{natbib}
\usepackage{subfigure}
\usepackage{float}
\usepackage{breqn}
\usepackage{float}
\usepackage{balance}
\usepackage{makecell}
\usepackage{bibspacing}
\usepackage{lipsum}
\usepackage{bm}

\usepackage[hang,flushmargin]{footmisc}  

\title{EmoDiff: Intensity Controllable Emotional Text-to-Speech with Soft-Label Guidance}
\name{Yiwei Guo, Chenpeng Du, Xie Chen, Kai Yu\sthanks{Corresponding author}}
\address{MoE Key Lab of Artificial Intelligence, AI Institute\\X-LANCE Lab, Department of Computer Science and Engineering\\Shanghai Jiao Tong University, Shanghai, China\\\texttt{\{cantabile\_kwok, duchenpeng, chenxie95, kai.yu\}@sjtu.edu.cn}}
\begin{document}
\ninept

\maketitle
\begin{abstract}
Although current neural text-to-speech (TTS) models are able to generate high-quality speech, intensity controllable emotional TTS is still a challenging task. 
Most existing methods need external optimizations for intensity calculation, leading to suboptimal results or degraded quality.
In this paper, we propose EmoDiff, a diffusion-based TTS model where emotion intensity can be manipulated by a proposed soft-label guidance technique derived from classifier guidance.
Specifically, instead of being guided with a one-hot vector for the specified emotion, EmoDiff is guided with a soft label where the value of the specified emotion and \textit{Neutral} is set to $\alpha$ and $1-\alpha$ respectively. 
The $\alpha$ here represents the emotion intensity and can be chosen from 0 to 1. 
Our experiments show that EmoDiff can precisely control the emotion intensity while maintaining high voice quality. Moreover, diverse speech with specified emotion intensity can be generated by sampling in the reverse denoising process. 

\end{abstract}
%
\section{Introduction}
\label{sec:intro}


Although current neural text-to-speech (TTS) models are able to
generate high-quality speech, such as Grad-TTS \cite{gradtts}, VITS \cite{vits} and VQTTS \cite{VQTTS}, intensity controllable emotional TTS
is still a challenging task. Unlike prosody modelling in recent literatures \cite{phone_level_3dim,du2021phone,guo2022unsupervised} that no specific label is provided in advance, emotional TTS typically utilizes dataset with categorical emotion labels.
Mainstream emotional TTS models \cite{tag1,tag2} can only synthesize emotional speech given the emotion label without intensity controllability.

In intensity controllable TTS models, efforts have been made to properly define and calculate emotion intensity values for training.
The most preferred method to define and obtain emotion intensity is the relative attributes rank (RAR)\cite{relative-attributes}, which is used in \cite{zhu2019controlling,lei2021fine,schnell2021improving,lei2022msemotts,MixedEmotion}.
RAR seeks a ranking matrix by a max-margin optimization problem, which is solved by support vector machines.
The solution is then fed to the model for training.
As this is a manually constructed and separated stage, it might result in suboptimal results that bring bias into training.
In addition to RAR, the operation on emotion embedding space is also explored.
\cite{um2020emotional} designs an algorithm to maximize distance between emotion embeddings, and interpolates the embedding space to control emotion intensity.
\cite{im2022emoq} quantizes the distance of emotion embeddings to obtain emotion intensities.
However, the structure of the embedding space also greatly influences the performance of these models, resulting in the need for careful extra constraints.
Intensity control for emotion conversion is investigated in \cite{choi2021sequence,Emovox}, with similar methods.
Some of the mentioned works also have degraded speech quality.
As an example, \cite{MixedEmotion} (which we refer to as ``MixedEmotion" later) is an autoregressive model with intensity values from RAR to weight the emotion embeddings.
It adopts pretraining to improve synthetic quality, but still with obvious quality degradation.

To overcome these issues, we need a conditional sampling method that can directly control emotions weighted with intensity.
In this work, we propose a soft-label guidance technique, based on the classifier guidance technique \cite{beat-gan,liu2019more} in denoising diffusion models \cite{DDPM,song-sde}.
Classifier guidance is an efficient sampling technique that uses the gradient of a classifier to guide the sampling trajectory given a one-hot class label.
In this paper, based on the extended soft-label guidance, we propose EmoDiff which is an emotional TTS model with sufficient intensity controllability.
Specifically, we first train an emotion-unconditional acoustic model. 
Then an emotion classifier is trained on any $\bm x_t$ on the diffusion process trajectory where $t$ is the diffusion timestamp.
In inference, we guide the reverse denoising process with the classifier and a soft emotion label where the value of the specified emotion and \textit{Neutral} is set to $\alpha$ and $1-\alpha$ respectively, instead of a one-hot distribution where only the specified emotion is 1 while all others are 0.
$\alpha\in[0,1]$ here represents the emotion intensity. 
Our experiments show that EmoDiff can precisely control the emotion intensity while maintaining high voice quality.
Moreover, it also generates diverse speech samples even with the same emotion as a strength of diffusion models \cite{beat-gan}.

In short words, the main advantages of EmoDiff are:
\begin{enumerate}
\item We define the emotion intensity as the weight for classifier guidance when using soft-labels. This achieves precise intensity control in terms of classifier probability, needless for extra optimizations. Thus it enables us to generate speech with arbitrary specified emotion intensity effectively.
\item It poses no harm to the synthesized speech. The generated samples have good quality and naturalness.
\item It also generates diverse samples even in the same emotion.
\end{enumerate}



\section{diffusion models with classifier guidance}
\label{sec:diffusion}
\subsection{Denoising Diffusion Models and TTS Applications}
Denoising diffusion probabilistic models \cite{DDPM,song-sde} have proven successful in many generative tasks. 
In the score-based interpretation \cite{song-score-matching,song-sde}, diffusion models construct a forward stochastic differential equation (SDE) to transform the data distribution $p_0(\bm x_0)$ into a known distribution $p_T(\bm x_T)$, and use a corresponding reverse-time SDE to generate realistic samples starting from noises. 
Thus, the reverse process is also called ``denoising" process.
Neural networks are then to estimate the score function $\nabla_ {\bm x} \log p_t(\bm x_t)$ for any $t\in[0,T]$ on the SDE trajectory, with score-matching objectives \cite{song-score-matching,song-sde}. 
In applications, 
diffusion models bypass the training instability and mode collapse problem in GANs, 
and outperform previous methods on sample quality and diversity \cite{beat-gan}.

Denoising diffusion models have also been used in TTS \cite{difftts,gradtts,diffsinger,fastdiff,lam2022bddm} and vocoding \cite{wavegrad,diffwave} tasks, with remarkable results.
In this paper, we build EmoDiff on the design of GradTTS \cite{gradtts}.
Denote $ \bm x\in \mathbb R^d$ a frame of mel-spectrogram, it constructs a forward SDE:
\begin{equation}\label{forward-SDE}
    \mathrm d \bm x_t = \frac 12  \bm\Sigma^{-1}(\bm\mu - \bm x_t)\beta_t\mathrm dt + \sqrt{\beta_t}\mathrm d \bm B_t
\end{equation}
where  $\bm B_t$ is a standard Brownian motion and $t\in[0,1]$ is the SDE time index.
$\beta_t$ is referred to as noise schedule such that $\beta_t$ is increasing and $\exp\left\{-\int _0^1 \beta_s\mathrm ds\right\}\approx 0$.
Then we have $p_1(\bm x_1)\approx \mathcal N(\bm x;\bm\mu,\bm \Sigma)$.
This SDE also indicates the conditional distribution $\bm x_t\mid \bm x_0\sim \mathcal N(\bm \rho(\bm x_0,\bm \Sigma,\bm \mu,t), \bm \lambda(\bm \Sigma,t))$, where $\bm \rho(\cdot),\bm \lambda (\cdot)$ both has closed forms.
Thus we can directly sample $\bm x_t$ from $\bm x_0$.
In practice, we set $\bm\Sigma$ to identity matrix and $\bm\lambda (\bm\Sigma,t)$ therefore becomes $\lambda_t\bm I$ where $\lambda_t$ is a scalar with known closed form.
Meanwhile, we condition the terminal distribution $p_1(\bm x_1)$ on text, i.e. let $\bm\mu=\bm\mu_\theta(\bm y)$, where $\bm y$ is the aligned phoneme representation of that frame.

The SDE of Eq.\eqref{forward-SDE} has a reverse-time counterpart:
\begin{equation}\label{reverse-time-SDE}
    \mathrm d\bm x_t =\left(\frac 12 \bm\Sigma^{-1}(\bm \mu-\bm x_t)-\nabla_{\bm x}\log p_t(\bm x_t)\right)\beta_t\mathrm dt + \sqrt{\beta_t}\mathrm d\bm {\widetilde{B}}_t
\end{equation}
where $\nabla \log p_t(\bm x_t)$ is the score function that is to be  estimated, and $\bm{\widetilde{B}}_t$ is a reverse-time Brownian motion.
It shares the trajectory of distribution $p_t(\bm x_t)$ with forward SDE in Eq.\eqref{forward-SDE}.
So, solving it from $\bm x_1\sim\mathcal N(\bm \mu, \bm\Sigma)$, we can end up with a realistic sample $\bm x_0\sim p(\bm x_0\mid \bm y)$. 
A neural network $\bm s_\theta(\bm x_t, \bm y, t)$ is trained to estimate the score function, in the following score-matching \cite{song-score-matching} objective:
\begin{equation}\label{eq:diff-loss}
    \min_\theta \mathcal L=\mathbb E_{\bm x_0,\bm y,t}[ \lambda_t\|\bm s_\theta(\bm x_t,\bm y,t)-\nabla_{\bm x_t}\log p(\bm x_t\mid \bm x_0)\|^2].
\end{equation}

\vspace{-.2in}
\subsection{Conditional Sampling Based on Classifier Guidance}

\begin{figure*}
    \centering
    \includegraphics[width=0.87\linewidth]{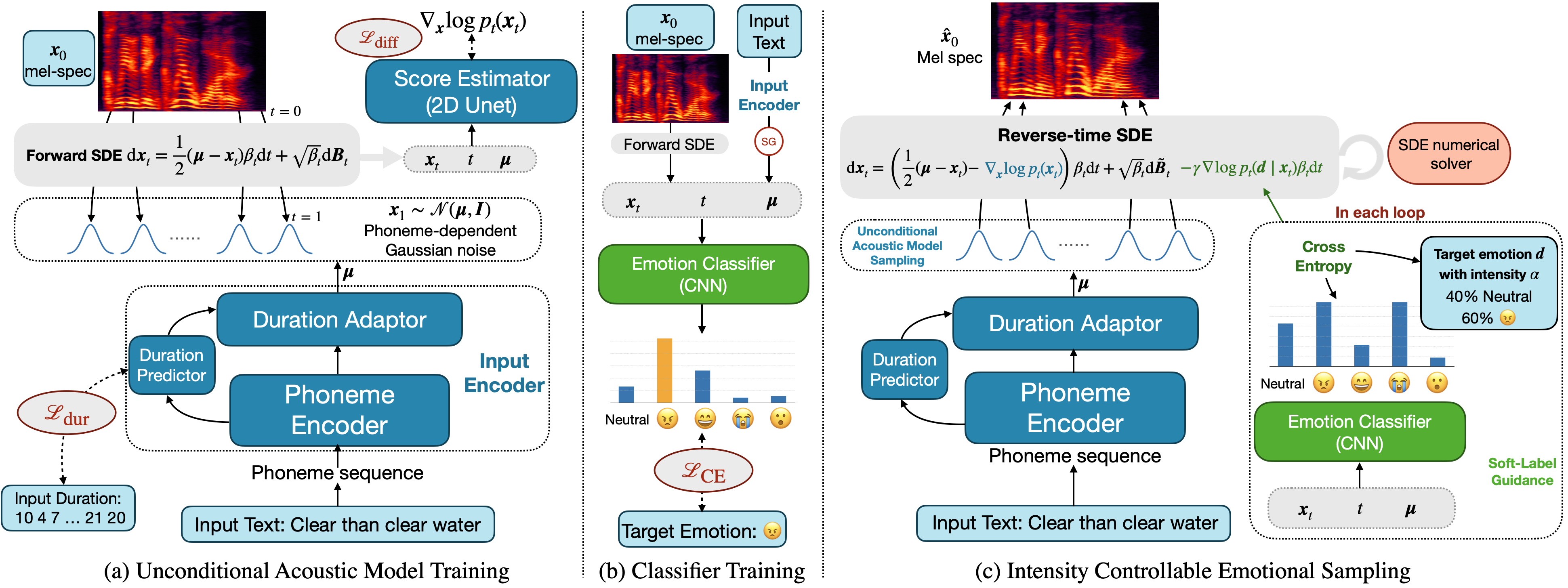}
    \caption{Training and sampling diagrams of EmoDiff. In training, $\bm x_t$ is directly sampled from known distribution $p(\bm x_t\mid \bm x_0)$. When sampling with a certain emotion intensity, the score function $\nabla_{\bm x} \log p_t(\bm x_t)$ is estimated by score estimator. ``SG" means stop gradient operation.}
    \label{fig:main}
\end{figure*}

\label{sec:classifier-guidance-basic}
Denoising diffusion models provide a new way of modeling conditional probabilities $p(\bm x \mid c)$ where $c$ is a class label.
Suppose we now have an unconditional generative model $p(\bm x)$, and a classifier $p(c\mid \bm x)$. 
By Bayes formula, we have
\begin{equation}\label{nabla-bayes}
    \nabla_{\bm x} \log p(\bm x\mid c) = \nabla_{\bm x} \log p(c\mid \bm x) + \nabla_{\bm x} \log p(\bm x).
\end{equation}
In the diffusion framework, to sample from conditional distribution $p(\bm x\mid c)$, we need to estimate score function $\nabla_{\bm x}\log p(\bm x_t\mid c)$.
By Eq.\eqref{nabla-bayes}, we only need to add the gradient from a classifier to the unconditional model.
This conditional sampling method is named classifier guidance \cite{beat-gan,liu2019more}, and is also used in unsupervised TTS \cite{guidedtts}.

In practice, classifier gradients are often scaled \cite{beat-gan,guidedtts} to control the strength of guidance.
Instead of original $\nabla_{\bm x} \log p(c\mid \bm x)$ in Eq.\eqref{nabla-bayes}, we now use $\gamma \nabla_{\bm x} \log p(c\mid \bm x)$, where $\gamma \geq  0$ is called guidance level. 
Larger $\gamma$ will result in highly class-correlated samples while smaller one will encourage sample variability \cite{beat-gan}.

Different from ordinary classifiers, the input to the classifier used here is all the $\bm x_t$ along the trajectory of SDE in Eq.\eqref{forward-SDE}, instead of clean $\bm x_0$ only. The time index $t$ can be anything in $[0, 1]$.
Thus, the classifier can also be denoted as $p(c\mid \bm x_t,t)$.

While Eq.\eqref{eq:nabla-bayes-extension} can effectively control sampling on class label $c$, it cannot be directly applied to soft-labels, i.e. labels weighted with intensity, as the guidance $p(c\mid \bm x)$ is not well-defined now.
Therefore, we extend this technique for emotion intensity control in Section \ref{sec:soft-label-guidance}.

\vspace{-.1in}
\section{EmoDiff}

\label{sec:EmoDiff}
\subsection{Unconditional Acoustic Model and Classifier Training}
The training of EmoDiff mainly includes the training of the unconditional acoustic model and emotion classifier. 
We first train a diffusion-based acoustic model on emotional data, but don't provide it with emotion conditions. 
This is referred to as ``unconditional acoustic model training" as in 
Figure \ref{fig:main}(a).
This model is based on GradTTS \cite{gradtts}, except that we provide explicit duration sequence by forced aligners to ease duration modeling.
In this stage, the training objective is $\mathcal L_{\text{dur}} + \mathcal L_{\text{diff}}$, where $\mathcal L_{\text{dur}}$ is the $\ell_2$ loss of logarithmic duration, and $\mathcal L_{\text{diff}}$ is the diffusion loss as Eq.\eqref{eq:diff-loss}.
In practice, following GradTTS, we also adopt prior loss $\mathcal L_{\text{prior}}=-\log \mathcal N(\bm x_0; \bm \mu,\bm I)$ to encourage converging. 
For notation simplicity, we use $\mathcal L_{\text{diff}}$ to denote diffusion and prior loss together in Figure \ref{fig:main}(a).

After training, the acoustic model can estimate score function of noisy mel-spectrogram $\bm x_t$ given input phoneme sequence $\bm y$, i.e. $\nabla\log p(\bm x_t\mid \bm y)$, which is unconditonal of emotion labels.  
Following Section \ref{sec:classifier-guidance-basic}, we then need an emotion classifier to distinguish emotion categories $e$ from noisy mel-spectrograms $\bm x_t$.
Meanwhile, as we always have a text condition $\bm y$, the classifier is formulated as $p( e \mid \bm x_t, \bm y, t)$.
As is shown in Figure \ref{fig:main}(b), 
the input to the classifier consists of three components: SDE timestamp $t$, noisy mel-spectrogram $\bm x_t$ and phoneme-dependent Gaussian mean $\bm \mu$. 
This classifier is trained with the standard cross-entropy loss $\mathcal L_{\text{CE}}$.
Note that we freeze the acoustic model parameters in this stage, and only update the weights in emotion classifier.

As we always need text $\bm y$ as condition along through the paper, we omit it and denote this classifier as $p(e\mid \bm x)$ in later sections to simplify the notation, if no ambiguity is caused.

\subsection{Intensity Controllable Sampling with Soft-Label Guidance}
\label{sec:soft-label-guidance}
In this section, we extend the classifier guidance to soft-label guidance which can control emotion weighted with intensity.
Suppose the number of basic emotions is $m$, and every basic emotion $e_i$ has a one-hot vector form $\bm e_i\in\mathbb R^m, i\in\{0,1,...,m-1\}$.
For each $\bm e_i$, only the $i$-th dimension is 1.
We specially use $\bm e_0$ to denote \textit{Neutral}.
For an emotion weighted with intensity $\alpha$ on $\bm e_i$, we define it to be $\bm d = \alpha \bm e_i + (1-\alpha) \bm e_0$. 
Then the gradient of log-probability of clasifier $p(\bm d\mid \bm x)$ w.r.t $\bm x$ can be defined as
\begin{equation}\label{eq:weighted-combination}
    \nabla_{\bm x} \log p(\bm d\mid \bm x) \triangleq \alpha \nabla_{\bm x} \log p( e_i\mid \bm x) + (1-\alpha)\nabla_{\bm x}\log p( e_0\mid \bm x).
\end{equation}
The intuition of this definition is that, intensity $\alpha$ stands for the contribution of emotion $e_i$ on the sampling trajectory of $\bm x$.
Larger $\alpha$ means we sample $\bm x$ along a trajectory with large ``force" towards emotion $e_i$, otherwise $e_0$.
Thus we can extend Eq.\eqref{nabla-bayes} to
\begin{align}\label{eq:nabla-bayes-extension}
    \nabla_{\bm x} \log p(\bm x\mid \bm d)= \alpha \nabla_{\bm x} \log p( e_i\mid \bm x)& + (1-\alpha)\nabla_{\bm x}\log p( e_0\mid \bm x)\nonumber \\ 
    &+ \nabla_{\bm x} \log p(\bm x).
\end{align}
When the intensity $\alpha$ is $1.0$ (100\% emotion $e_i$) or $0.0$ (100\% \textit{Neutral}), the above operation reduces to the standard classifier guidance form Eq.\eqref{nabla-bayes}.
Hence we can use the soft-label guidance Eq.\eqref{eq:weighted-combination} in the sampling process, and generate a realistic sample with specified emotion $\bm d=\alpha \bm e_i + (1-\alpha)\bm e_0$ with intensity $\alpha$.

Figure \ref{fig:main}(c) illustrates the intensity controllable sampling process. 
After feeding the acoustic model and obtaining phoneme-dependent $\bm \mu$ sequence, we sample $\bm x_1\sim\mathcal N(\bm\mu, \bm I)$ and simulate reverse-time SDE from $t=1$ to $t=0$ through a numerical simulator.
In each simulator update, we feed the classifier with current $\bm x_t$ and get the output probabilities $p_t(\cdot \mid \bm x_t)$.
Eq.\eqref{eq:nabla-bayes-extension} is then used to calculate the guidance term.
Similar as Section \ref{sec:classifier-guidance-basic}, we also scale the guidance term with guidance level $\gamma$.
At the end, we obtain $\hat {\bm x}_0$ which is not only intelligible with input text, but also corresponding to the target emotion $\bm d$ with intensity $\alpha$.
This lead to precise intensity that correlates well to classifier probability.

Generally, in addition to intensity control, our soft-label guidance is capable for more complicated control on mixed emotions \cite{MixedEmotion}.
Denote $\bm d=\sum_{i=0}^{m-1}w_i\bm e_i$ a combination of all emotions where $w_i\in[0,1],\sum_{i=0}^{m-1}w_i=1$, 
Eq.\eqref{eq:weighted-combination} can be generalized to 
\begin{equation}\label{eq:general-weighted-sum}\vspace{-0.03in}  
    \nabla_{\bm x}\log p(\bm d\mid\bm x)\triangleq \sum_{i=0}^{m-1} w_i \nabla_{\bm x}\log p(e_i\mid \bm x).\vspace{-.02in}
\end{equation} 
Then Eq.\eqref{eq:nabla-bayes-extension} can also be expressed in such generalized form.
This extension can also be interpreted from the probabilistic view.
As the combination weights $\{w_i\}$ can be viewed as a categorical distribution $p_e(\cdot)$ over basic emotions $\{e_i\}$,
Eq.\eqref{eq:general-weighted-sum} is equivalent to
\begin{align}
    \nabla_{\bm x} \log p(\bm d\mid \bm x) &\triangleq \mathbb E_{e\sim p_e} \nabla_{\bm x} \log p(e\mid \bm x) \\
    &= -\nabla_{\bm x}  \operatorname{CE}\left[p_e(\cdot), p(\cdot \mid \bm x) \right]\label{eq:cross-entropy-derivation}
\end{align}
where $\operatorname{CE}$ is the cross-entropy function.
Eq.\eqref{eq:cross-entropy-derivation} implies the fact that we are actually decreasing the cross-entropy of target emotion distribution $p_e$ and classifier output $p(\cdot \mid \bm x)$, when sampling along the gradient $\nabla_{\bm x} \log p(\bm d\mid \bm x)$.
The gradient of cross-entropy w.r.t $\bm x$ can guide the sampling process.
Hence, this soft-label guidance technique can generally be used to control any arbitrary complex emotion as a weighted combination of several basic emotions.

In Figure \ref{fig:main}(c), we use cross-entropy as a concise notation for soft-label guidance term. In our intensity control scenario, it reduces to Eq.\eqref{eq:weighted-combination} mentioned before.

\vspace{-.1in}
\section{Experiments and Results}
\begin{figure*}
    \centering
    \includegraphics[width=0.99\linewidth]{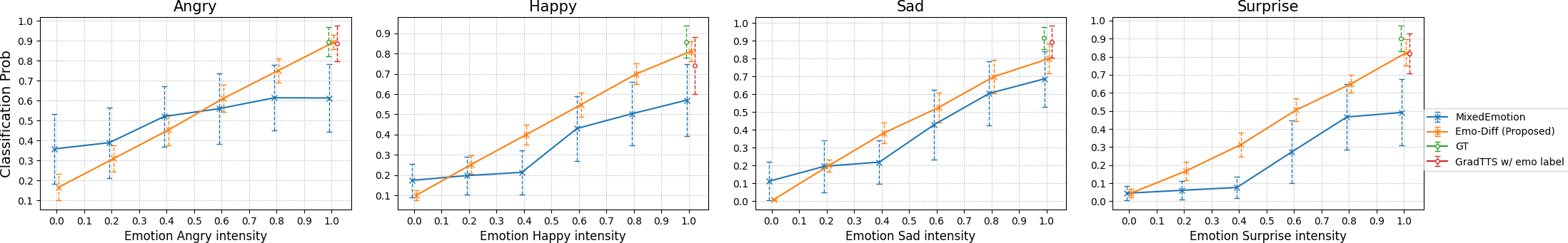}
    \caption{Classification probabilities when controlling on intensity $\alpha\in\{0.0,0.2,0.4,0.6,0.8,1.0\}$. Errorbars represent standard deviation.}
    \label{fig:classifier-objective}
\end{figure*}

\label{sec:exp}

\subsection{Experimental Setup}
We used the English part of the Emotional Speech Dataset (ESD) \cite{zhoukun-ESD} to perform all the experiments.
It has 10 speakers, each with 4 emotional categories \textit{Angry, Happy, Sad, Surprise} together with a \textit{Neutral} category. There are 350 parallel utterances per speaker and emotion category, amounting to about 1.2 hours each speaker. 
Mel-spectrogram and forced alignments were extracted by Kaldi \cite{povey2011kaldi} in 12.5ms frame shift and 50ms frame length, followed by cepstral normalization. 
Audio samples in these experiments are available \footnote{\scriptsize{\url{https://cantabile-kwok.github.io/EmoDiff-intensity-ctrl/}}}.

In this paper, we only consider single-speaker emotional TTS problem.
Throughout the following sections, we trained an unconditional GradTTS acoustic model on all 10 English speakers for a reasonable data coverage, and a classifier on a female speaker (ID:0015) only.
The unconditional GradTTS model was trained with Adam optimizer at $10^{-4}$ learning rate for 11M steps.
We used exponential moving average on model weights as it is reported to improve diffusion model's performance \cite{song-sde}.
The structure of the classifier is a 4-layer 1D CNN, with BatchNorm and Dropout in each block.
In the inference stage, guidance level $\gamma$ was fixed to 100.

We chose HifiGAN \cite{hifigan} trained on all the English speakers here as a vocoder for all the following experiments.

\vspace{-.1in}
\subsection{Emotional TTS Quality}
\begin{table}[t]
\caption{
MOS and MCD comparisons.
MOS is presented with 95\% confidence interval.
Note that  ``GradTTS w/ emo label" cannot control emotion intensity.
}
\centering
\begin{tabular}{lcc}
\toprule
 & \textbf{MOS} & \textbf{MCD$_{25}$} \\
 \midrule
GT & 4.73$\pm$0.09  & - \\
GT (voc.) &4.69$\pm$0.10  & 2.96 \\ \hline
MixedEmotion \cite{MixedEmotion} & 3.43$\pm$0.12 &  6.62\\
GradTTS w/ emo label & 4.16$\pm$0.10 & 5.75\\ \hline
EmoDiff (ours) & 4.13$\pm$0.10 & 5.98\\
\bottomrule
\end{tabular}
\label{tab:MOS-MCD}
\vspace{-.1in}
\end{table}

We first measure the speech quality, which contains audio quality and speech naturalness.
We did comparisons of the proposed EmoDiff with the following systems:
\begin{enumerate}
    \item \textbf{GT} and \textbf{GT (voc.)}: ground truth recording and analysis synthesis result (vocoded with GT mel-spectrogram).
    \item \textbf{MixedEmotion}\footnote{We used the official implementation \url{https://github.com/KunZhou9646/Mixed_Emotions}}: proposed in \cite{MixedEmotion}. It is an autoregressive model based on relative attributes rank to pre-calculate intensity values for training. 
    It much resembles Emovox \cite{Emovox} for intensity controllable emotion conversion.
    \item \textbf{GradTTS w/ emo label}: a conditional GradTTS model with hard emotion labels as input. 
    It therefore does not have intensity controllability, but should have good sample quality, as a certified acoustic model.
\end{enumerate}
Note that in this experiment, samples from EmoDiff and MixedEmotion were controlled with $\alpha = 1.0$ intensity weight, so that they are directly comparable with others.

Table \ref{tab:MOS-MCD} presents the mean opinion score (MOS) and mel cepstral distortion (MCD) evaluations.
It is shown that the vocoder causes little deterioration on sample quality, and our EmoDiff outperforms MixedEmotion baseline with a large margin.
Meanwhile, EmoDiff and the hard-conditioned GradTTS both have decent and very close MOS results.
The MCD results of them only have a small difference.
This means EmoDiff does not harm sample quality for intensity controllability, unlike MixedEmotion.

\subsection{Controllability of Emotion Intensity}

To evaluate the controllability of emotion intensity, we used our trained classifier to classify the synthesized samples under a certain intensity that was being controlled.
The $t$ input to the classifier was now set to $0$.
The average classification probability on the target emotion class was used as the evaluation metric.
Larger values indicate large discriminative confidence.
For both EmoDiff and MixedEmotion on each emotion, we varied the intensity from $\alpha=0.0$ to $1.0$.
When intensity is $0.0$, it equivalents to synthesize 100\% \textit{Neutral} samples.
Larger intensity should result in larger  probability.

Figure \ref{fig:classifier-objective} presents the results.
To demonstrate the capability of this classifier, we plotted the classification probability on ground truth data.
To show the performance of hard-conditioned GradTTS model, we also plotted the probability on its synthesized samples.
As it doesn't have intensity controllability, we only plotted the values when intensity was $1.0$.
Standard deviations are presented as an errobar here as well for each experiment.

It can be found from the figure that the trained classifier has a reasonable performance on ground truth data at first.
As a remark, the classification accuracy on validation set is 93.1\%.
Samples from GradTTS w/ emo label have some lower classification probabilities.
Most importantly, the proposed EmoDiff always covers a larger range from intensity $\alpha=0.0$ to $1.0$ than the baseline.
The error range of EmoDiff is also always lower than the baseline, meaning that our control is more stable.
This proves the effectiveness of our proposed soft-label guidance technique.
We also notice that sometimes EmoDiff reaches higher classification probability than hard-conditioned GradTTS at intensity $1.0$.
This is also reasonable, as conditioning on emotion labels when training is not guaranteed to achieve better class-correlation than classifier guidance, with a strong classifier and sufficient guidance level.

\vspace{-.1in}
\subsection{Diversity of Emotional Samples}

\begin{figure}
    \centering
    \includegraphics[width=0.8\linewidth]{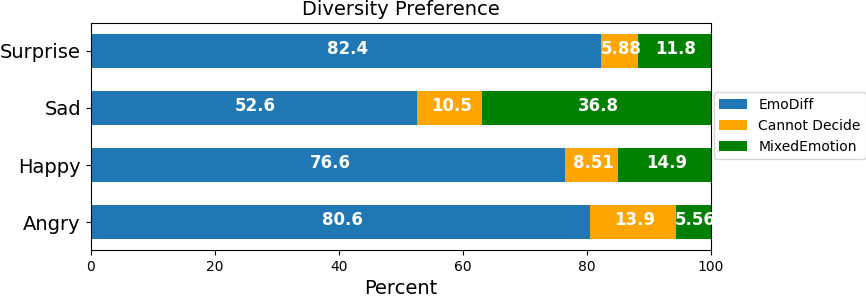}
    \caption{Diversity preference test of each emotion.}
    \label{fig:diversity}
    \vspace{-.2in}
\end{figure}

Despite genearating high-quality and intensity controllable emotional samples, EmoDiff also has good sample diversity even in the same emotion, benefiting from the powerful generative ability of diffusion models.
To evaluate the diversity of emotional samples, we conducted a subjective preference test for each emotion between our EmoDiff and MixedEmotion.
Listeners were asked to choose the more diverse one, or ``Cannot Decide".
Note that the test was done for each emotion in $\alpha = 1.0$ weight.

Figure \ref{fig:diversity} shows the preference result.
It is clear that for each of the three emotion categories \textit{Angry, Happy} and \textit{Surprise}, EmoDiff owns a large advantage of being preferred in diversity.
Only for \textit{Sad}, EmoDiff outperforms the baseline with a little margin.
This is mainly because MixedEmotion is autoregressive, and we found its variation on duration accounts much especially for \textit{Sad} samples.

\vspace{-.1in}
\section{Conclusion}
In this paper, we investigated the intensity control problem in emotional TTS systems.
We defined emotion with intensity to be the weighted sum of a specific emotion and \textit{Neutral}, with the weight being the intensity values.
Under this modeling, we extended classifier guidance technique to soft-label guidance, which enables us to directly control any arbitrary emotion with intensity instead of a one-hot class label.
By this technique, the proposed EmoDiff can achieve simple but effective control on emotion intensity, with an unconditional acoustic model and emotion classifier.
Subjective and objective evaluations demonstrated that EmoDiff outperforms baseline in terms of TTS quality, intensity controllability and sample diversity.
Also, the proposed soft-label guidance can generally be applied to control more complicated natural emotions, which we leave as a future work.
\label{sec:conclusion}

\vspace{-0.06in}
\section{Acknowledgements}
\vspace{-0.04in}
This study is supported by Shanghai Municipal Science and Technology Major Project
(2021SHZDZX0102), Jiangsu Technology Project (No.BE2022059-2).

\newpage

\bibliographystyle{IEEEtran}
\bibliography{strings,refs}
\end{document}